\documentclass[smallextended]{svjour3}
\usepackage{graphicx, amsmath, amssymb}
\usepackage[caption=false]{subfig}

\newcommand{\ket}[1]{{\left| #1 \right\rangle}}
\newcommand{\ketbra}[2]{{\left| #1 \middle\rangle \middle \langle #2 \right|}}

\journalname{Quantum Inf Process}

\begin{document}

\title{Laplacian versus Adjacency Matrix in Quantum Walk Search}

\author{Thomas G.~Wong \and Lu\'{i}s Tarrataca \and Nikolay Nahimov}

\authorrunning{T.~G.~Wong, L.~Tarrataca, N.~Nahimov}

\institute{T.~G.~Wong \at
	   Faculty of Computing, University of Latvia, Rai\c{n}a bulv.~19, R\=\i ga, LV-1586, Latvia \\
	   \email{twong@lu.lv}
	   \and
	   L.~Tarrataca \at
	   Laborat\'{o}rio Nacional de Computa\c{c}\~{a}o Cient\'{\i}fica, Petr\'{o}polis, Brazil \\
	   \email{luis.tarrataca@gmail.com}
	   \and
	   N.~Nahimov \at
	   Faculty of Computing, University of Latvia, Rai\c{n}a bulv.~19, R\=\i ga, LV-1586, Latvia \\
	   \email{nikolajs.nahimovs@lu.lv}
}

\date{Received: date / Accepted: date}

\maketitle

\begin{abstract}
	A quantum particle evolving by Schr\"{o}dinger's equation contains, from the kinetic energy of the particle, a term in its Hamiltonian proportional to Laplace's operator. In discrete space, this is replaced by the discrete or graph Laplacian, which gives rise to a continuous-time quantum walk. Besides this natural definition, some quantum walk algorithms instead use the adjacency matrix to effect the walk. While this is equivalent to the Laplacian for regular graphs, it is different for non-regular graphs, and is thus an inequivalent quantum walk. We algorithmically explore this distinction by analyzing search on the complete bipartite graph with multiple marked vertices, using both the Laplacian and adjacency matrix. The two walks differ qualitatively and quantitatively in their required jumping rate, runtime, sampling of marked vertices, and in what constitutes a natural initial state. Thus the choice of the Laplacian or adjacency matrix to effect the walk has important algorithmic consequences.
\end{abstract}


\section{Introduction}

Schr\"odinger's equation \cite{Schrodinger1926}
\[ i \frac{\partial \psi}{\partial t} = H \psi \]
is the fundamental equation in quantum mechanics \cite{Griffiths2005}, describing the time-evolution of a quantum state $\psi$ generated by Hamiltonian $H$. Note we have set $\hbar = 1$. The Hamiltonian characterizes the total energy of the system, and for a particle of mass $m$, it includes a kinetic energy term
\[ \frac{-1}{2m} \nabla^2, \]
where $\nabla^2 = \partial^2/\partial x^2 + \partial^2/\partial y^2 + \partial^2/\partial z^2$ is Laplace's operator (in three-dimensional Euclidean space).

If the particle is confined to discrete spatial locations, such as when a particle trapped in an optical lattice \cite{Bloch2005}, then $\nabla^2$ is replaced by the discrete or graph Laplacian $L = A - D$, where $A$ is the adjacency matrix ($A_{ij} = 1$ if $i$ and $j$ are connected, and $0$ otherwise), and $D$ is the diagonal degree matrix ($D_{ii} = {\rm deg}(i)$). For example, for a one-dimensional grid with lattice spacing $h$, note the similarities between the continuous-space Laplacian
\[ \nabla^2 \psi = \frac{d^2 \psi}{dx^2} = \lim_{h \to 0} \frac{\psi(x+h) + \psi(x-h) - 2 \psi(x)}{h^2} \]
and the discrete-space analogue
\[ L \psi = (A - D) \psi = \psi_{x+1} + \psi_{x-1} - 2 \psi_x. \]
Now letting $\gamma = 1/2m$, the kinetic energy operator becomes
\[ -\gamma L. \]
This defines a continuous-time quantum walk \cite{FG1998,CG2004}, and it is the natural movement of a quantum particle with kinetic energy when confined to a lattice. The real parameter $\gamma > 0$ corresponds to the jumping rate, or amplitude per time, of the walk. A higher jumping rate corresponds to a particle with less mass, since a less massive particle scatters more readily.

Besides this natural definition, any Hermitian operator (so that the time-evolution operator $e^{-iHt}$ is unitary) that respects the locality of the graph defines a continuous-time quantum walk. Another commonly used definition is the adjacency matrix $A$ \cite{Childs2003,FGG2008}, which differs from the Laplacian by dropping the degree matrix $D$. That is, the term in the Hamiltonian effecting the walk is
\[ -\gamma A. \]

These two common generators of the quantum walk, the Laplacian and adjacency matrix, can also arise in interacting spin models in statistical physics. In particular, a single excitation in a network of spins can be expressed as a particular spin being spin-up while the others are spin-down. With $XY\!Z$ interactions between nearest-neighbor spins (\textit{i.e.}, the Heisenberg model), the Hamiltonian reduces to the Laplacian of the graph, and with $XY$ interactions alone, the adjacency matrix arises instead \cite{Bose2009}.

If the graph is regular, then the degree matrix $D$ is a multiple of the identity matrix, so it only constitutes an unobservable, global phase or a rezeroing of energy and can be dropped \cite{CG2004}. Thus for regular graphs, the Laplacian and adjacency matrix are equivalent definitions of the quantum walk. If the graph is non-regular, however, then the two walks are inequivalent, and their differences have been explored for state transfer \cite{Alvir2015,Ackelsberg2015}.

\begin{figure}
\begin{center}
	\includegraphics{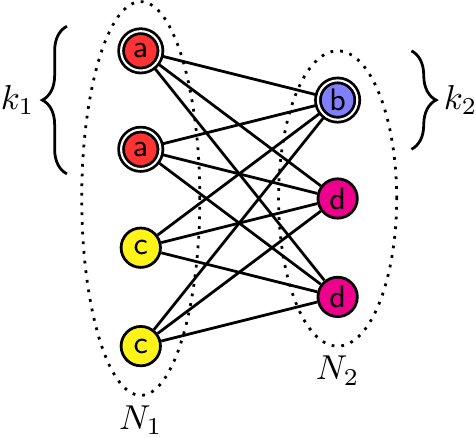}
	\caption{\label{fig:bipartite} Complete bipartite graph with $N_1 = 4$ and $N_2 = 3$ vertices in each vertex set, $k_1 = 2$ and $k_2 = 1$ of which are marked (denoted by double circles) in the respective sets. Identically evolving vertices are identically colored and labeled.}
\end{center}
\end{figure}

In this paper, we investigate the \emph{algorithmic} consequences of choosing the Laplacian or the adjacency matrix to effect the quantum walk by examining spatial search on the complete bipartite graph with multiple marked vertices, an example of which is shown in Fig.~\ref{fig:bipartite}. We take the number of vertices in each vertex set $V_1$ and $V_2$ to be $N_1$ and $N_2$ (so that the total number of vertices is $N = N_1 + N_2$), and since they are allowed to differ, the graph is, in general, non-regular. Thus the Laplacian and adjacency matrix effect different walks, and we expect the search algorithms to also behave differently. For simplicity, we assume that the size of each vertex set scales with the number of vertices (\textit{i.e.}, $N_1 = \Theta(N)$ and $N_2 = \Theta(N)$), and the number of marked vertices $k_1$ and $k_2$ in each set scales smaller (\textit{i.e.}, $k_1 = o(N)$ and $k_2 = o(N)$). This assumption avoids scenarios where a vertex set is small and can be classically brute-force searched in little time, or where the number of marked vertices in one set is larger than the number of vertices in the other set, thereby simplifying the analysis.

This investigation differs from previous work on spatial search by continuous-time quantum walk. For example, much of the existing literature focuses on search on regular graphs, for which the Laplacian and adjacency matrix are equivalent. This includes the complete graph \cite{CG2004,Wong10}, strongly regular graphs \cite{Wong5}, the hypercube \cite{CG2004}, arbitrary-dimensional periodic square lattices \cite{CG2004}, and the simplex of complete graphs \cite{Wong7,Wong9,Wong11,Wong16}. Even though the joined complete graphs in \cite{Wong7} form a non-regular graph, it is approximately regular, and so difference between the Laplacian and adjacency matrix is negligible. For spatial search on a truly non-regular graph, the work of \cite{Novo2015} is of note, where they investigate search on the complete bipartite graph with one marked vertex. They only considered the adjacency matrix, however, whereas our analysis includes multiple marked vertices and the Laplacian as well. Furthermore, their algorithm reaches a success probability of $1/2$ because, as we will show, a sub-ideal initial state is used; with the ideal initial state, the algorithm searches with certainty. Finally, previous work on search on Erd\"os-Renyi random graphs \cite{Chakraborty2015} was also restricted to quantum walks effected by the adjacency matrix alone. Our work here seems to be the first direct comparison between the Laplacian and adjacency matrix in spatial search by continuous-time quantum walk.

In the next section, we analyze search on the complete bipartite graph with multiple marked vertices (\textit{e.g.}, Fig.~\ref{fig:bipartite}) when the quantum walk is effected by the Laplacian. Depending on the choice of the jumping rate $\gamma$, the algorithm either finds the marked vertices in one vertex set or the other, and with different runtimes. Following this, we solve the search problem when the walk is governed by the adjacency matrix. What constitutes a natural initial state differs from the usual equal superposition, and with a particular choice of $\gamma$, the system evolves to a combination of all marked vertices, regardless of which vertex set they are in. Qualitatively, this is a different behavior, and the runtime is also quantitatively different.


\section{Laplacian Walk}

We begin with the quantum walk being generated by the Laplacian. In the standard spatial search algorithm by continuous-time quantum walk \cite{CG2004}, the system $\ket{\psi(t)}$ starts in an equal superposition $\ket{s}$ over the vertices:
\[ \ket{\psi(0)} = \ket{s} = \frac{1}{\sqrt{N}} \sum_{i = 0}^{N-1} \ket{i}. \]
This state expresses our initial lack of knowledge of where the marked vertices are, so it guesses each vertex with equal probability. Moreover, since it is an eigenvector of the Laplacian $L$ with eigenvalue $0$, evolving by the quantum walk alone ($H = -\gamma L$) causes the system to stay in $\ket{s}$, expressing our continued lack of information as to where the marked vertices might be.

With an oracle that identifies the marked vertices, however, our information does change, and the system evolves from $\ket{s}$. In particular, the search Hamiltonian is
\[ H = -\gamma L - \!\!\!\!\!\! \sum_{i \in {\rm marked}} \!\!\!\!\! \ketbra{i}{i}. \]
With this initial state and evolution, the system evolves in a four-dimensional (4D) subspace as shown in Fig.~\ref{fig:bipartite}. Grouping identically-evolving vertices together, we get a basis for the 4D subspace:
\begin{gather*}
	\ket{a} = \frac{1}{\sqrt{k_1}} \!\!\!\! \sum_{i \in V_1 \atop i \in {\rm marked}} \!\!\!\! \ket{i}, \quad
	\ket{b} = \frac{1}{\sqrt{k_2}} \!\!\!\! \sum_{i \in V_2 \atop i \in {\rm marked}} \!\!\!\! \ket{i}, \\
	\ket{c} = \frac{1}{\sqrt{N_1 - k_1}} \!\!\!\! \sum_{i \in V_1 \atop i \not\in {\rm marked}} \!\!\!\! \ket{i}, \quad
	\ket{d} = \frac{1}{\sqrt{N_2 - k_2}} \!\!\!\! \sum_{i \in V_2 \atop i \not\in {\rm marked}} \!\!\!\! \ket{i} .
\end{gather*}
In this $\{ \ket{a}, \ket{b}, \ket{c}, \ket{d} \}$ basis, the initial equal superposition state is
\[ \ket{s} = \frac{1}{\sqrt{N}} \begin{pmatrix}
	\sqrt{k_1} \\
	\sqrt{k_2} \\
	\sqrt{N_1-k_1} \\
	\sqrt{N_2-k_2} \\
\end{pmatrix}. \]
Additionally, the adjacency matrix in this basis is
\begin{equation}
	\label{eq:A}
	A = \begin{pmatrix}
		0 & \sqrt{k_1k_2} & 0 & \sqrt{k_1 N_{k2}} \\
		\sqrt{k_1k_2} & 0 & \sqrt{k_2 N_{k1}} & 0 \\
		0 & \sqrt{k_2 N_{k1}} & 0 & \sqrt{N_{k1} N_{k2}} \\
		\sqrt{k_1 N_{k2}} & 0 & \sqrt{N_{k1} N_{k2}} & 0 \\
	\end{pmatrix},
\end{equation}
where $N_{ki} = N_i - k_i$, and the degree matrix is
\[ D = \begin{pmatrix}
	N_2 & 0 & 0 & 0 \\
	0 & N_1 & 0 & 0 \\
	0 & 0 & N_2 & 0 \\
	0 & 0 & 0 & N_1 \\
\end{pmatrix}. \]
The Laplacian is simply $L = A - D$, so the search Hamiltonian is
\begin{equation}
	\label{eq:H}
	H = -\gamma \begin{pmatrix}
		\frac{1}{\gamma}-N_2 & \sqrt{k_1k_2} & 0 & \sqrt{k_1 N_{k2}} \\
		\sqrt{k_1k_2} & \frac{1}{\gamma}-N_1 & \sqrt{k_2 N_{k1}} & 0 \\
		0 & \sqrt{k_2 N_{k1}} & -N_2 & \sqrt{N_{k1} N_{k2}} \\
		\sqrt{k_1 N_{k2}} & 0 & \sqrt{N_{k1} N_{k2}} & -N_1 \\
	\end{pmatrix}.
\end{equation}

\begin{figure}
\begin{center}
	\includegraphics{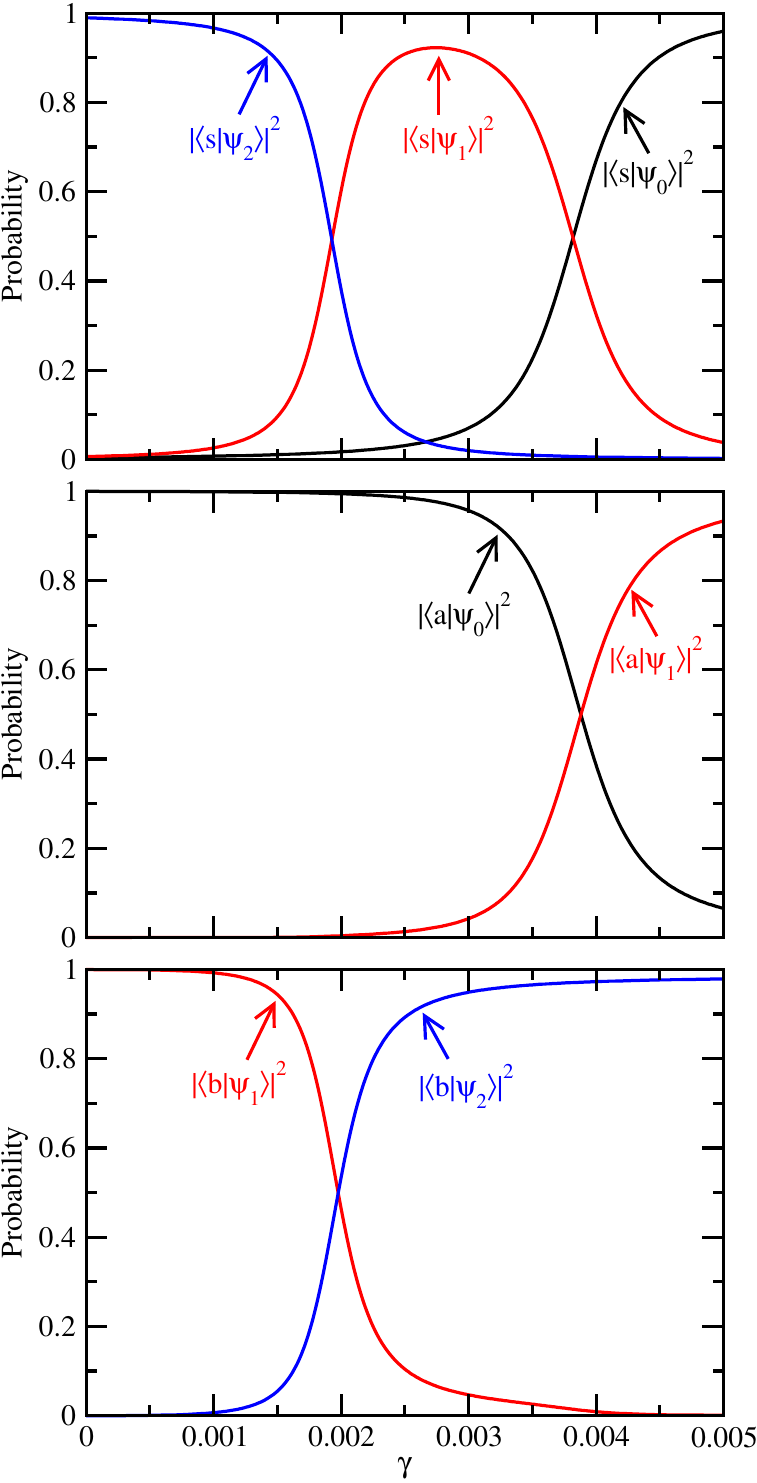}
	\caption{\label{fig:overlap_L} Probability overlaps of $\ket{s}$, $\ket{a}$, and $\ket{b}$ with eigenstates of $H$ for search on the complete bipartite graph with $N_1 = 512$, $N_2 = 256$, $k_1 = 3$, and $k_2 = 5$, where the walk is effected by the Laplacian.}
\end{center}
\end{figure}

In Fig.~\ref{fig:overlap_L}, we plot the probability overlaps of the eigenstates of $H$ with $\ket{s}$, $\ket{a}$, and $\ket{b}$, where $\ket{\psi_0}$ denotes the eigenvector of $H$ with the smallest eigenvalue, $\ket{\psi_1}$ denotes the eigenvector with the next smallest eigenvalue, and so forth. Since these eigenvalues correspond to the energy levels of the system \cite{Griffiths2005}, $\ket{\psi_0}$ is called the ground state and $\ket{\psi_{i > 0}}$ is called the $i$th excited state. This reveals that the behavior of the algorithm strongly depends on the value of $\gamma$. In particular, when $\gamma$ is small, the initial equal superposition state $\ket{s}$ is approximately the second excited state $\ket{\psi_2}$ of $H$, meaning the system starts in an eigenstate and fails to evolve apart from an unobservable, global phase. Similarly, when $\gamma$ takes intermediate or large values, $\ket{s}$ is approximately the first excited state $\ket{\psi_1}$ or ground state $\ket{\psi_0}$ of $H$, so again the system fails to evolve apart from an unobservable, global phase. It is only when $\gamma$ is near one of its two ``critical values'' \cite{CG2004}, where $\ket{s}$ exhibits a ``phase transition'' in which eigenstates of $H$ support it, that the system evolves substantially. In Fig.~\ref{fig:overlap_L}, the critical $\gamma$'s correspond to the crossings near $\gamma = 0.004$ and $\gamma = 0.002$. Let us respectively call these $\gamma_a$ and $\gamma_b$. When $\gamma = \gamma_a = 0.004$, both $\ket{\psi_0}$ and $\ket{\psi_1}$ are half $\ket{s}$ and half $\ket{a}$. This causes the system to evolve from $\ket{s}$ to $\ket{a}$ in time $\pi/\Delta E$, where $\Delta E$ is the energy gap between $\ket{\psi_1}$ and $\ket{\psi_0}$ \cite{CG2004,Wong10}. Similarly, when $\gamma = \gamma_b = 0.002$, both $\ket{s}$ and $\ket{b}$ are roughly half in $\ket{\psi_1}$ and half in $\ket{\psi_2}$, so the system evolves from $\ket{s}$ to $\ket{b}$ in time $\pi/\Delta E$, where $\Delta E$ is the energy gap between $\ket{\psi_2}$ and $\ket{\psi_1}$.

\begin{figure}
\begin{center}
	\subfloat[]{
		\includegraphics{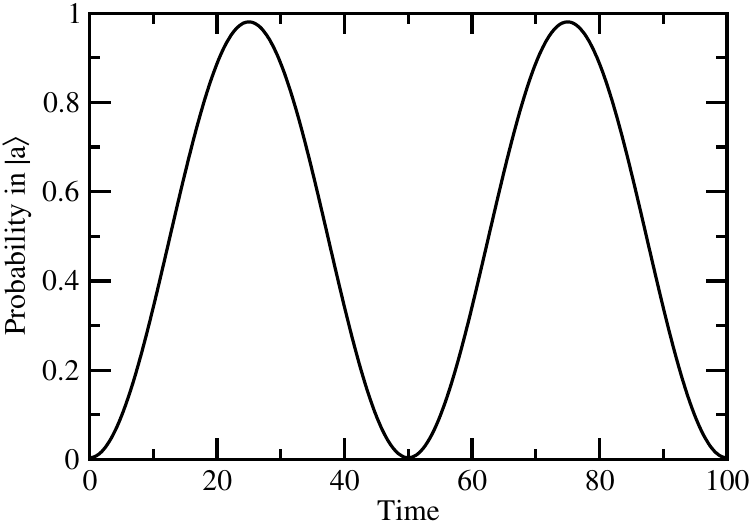}
		\label{fig:evolution_L_a}
	}

	\subfloat[]{
		\includegraphics{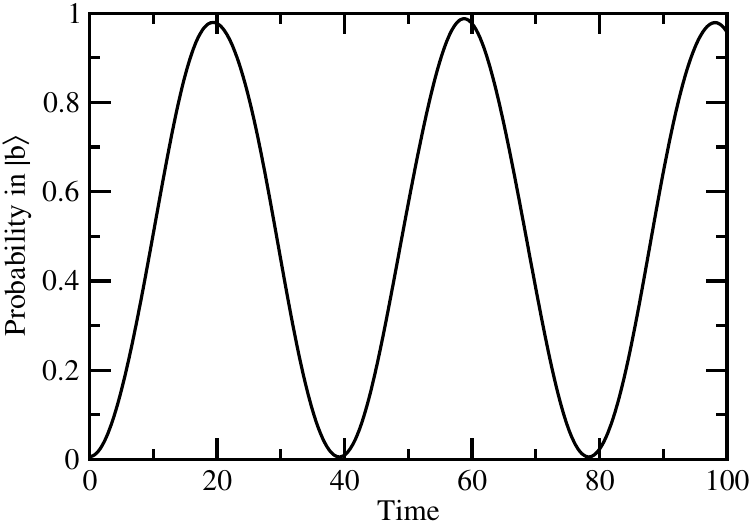}
		\label{fig:evolution_L_b}
	}
	\caption{For search on the complete bipartite graph with $N_1 = 512$, $N_2 = 256$, $k_1 = 3$, and $k_2 = 5$, where the walk is effected by the Laplacian: the success probability in (a) $\ket{a}$ when $\gamma = \gamma_a$ and (b) $\ket{b}$ when $\gamma = \gamma_b$.}
\end{center}
\end{figure}

Proving this simply involves finding the eigenvectors and eigenvalues of the search Hamiltonian \eqref{eq:H} when $\gamma$ takes its critical values of $\gamma_a$ and $\gamma_b$. We find these eigenvectors and eigenvalues in Appendix~\ref{appendix:Laplacian} using degenerate perturbation theory \cite{Wong5}, showing that when $\gamma$ is within $o(1/N^{3/2})$ of
\[ \gamma_a = \frac{1}{N_2},\]
then two of the (unnormalized) eigenvectors and corresponding eigenvalues of the search Hamiltonian \eqref{eq:H}, for large $N$, are
\[ \ket{s} \pm \ket{a}, \quad E = \mp \sqrt{\frac{k_1}{N}}. \]
Thus for large $N$, the system evolves from $\ket{s}$ to $\ket{a}$ in time
\[ t_a = \frac{\pi}{\Delta E} = \frac{\pi}{2} \sqrt{\frac{N}{k_1}}. \]
As a check, in Fig.~\ref{fig:evolution_L_a}, we plot the probability in $\ket{a}$ as the system evolves with $\gamma = \gamma_a = 1/256 \approx 0.004$. As expected, the success probability nears $1$ (with the slight deficiency from $1$ remedied by increasing $N$) at time $t_a = (\pi/2) \sqrt{(512 + 256)/3} = 8\pi \approx 25.133$.

Also shown in Appendix~\ref{appendix:Laplacian}, when $\gamma$ is within $o(1/N^{3/2})$ of the critical value of 
\[ \gamma_b = \frac{1}{N_1},\]
the search Hamiltonian \eqref{eq:H} for large $N$ has (unnormalized) eigenvectors and corresponding eigenvalues
\[ \ket{s} \pm \ket{b}, \quad E = \mp \sqrt{\frac{k_2}{N}}. \]
Thus for large $N$, the system evolves from $\ket{s}$ to $\ket{b}$ in time
\[ t_b = \frac{\pi}{\Delta E} = \frac{\pi}{2} \sqrt{\frac{N}{k_2}}. \]
As a check, in Fig.~\ref{fig:evolution_L_b}, we plot the probability in $\ket{b}$ as the system evolves with $\gamma = \gamma_b = 1/512 \approx 0.002$. As expected, the success probability nears $1$ at time $t_b = (\pi/2) \sqrt{(512 + 256)/5} \approx 19.468$.

Thus depending on whether $\gamma$ equals $\gamma_a$ or $\gamma_b$, the system evolves to one of $\ket{a}$ or $\ket{b}$, not a combination of $\ket{a}$ and $\ket{b}$. That is, the final state samples entirely from the marked vertices in one vertex set or the other. As we will see in the next section, this differs from search with the adjacency matrix, which samples from both vertex sets simultaneously. Of course, this excludes the special case when $N_1 = N_2 = N/2$, which causes the graph to be regular. Then the Laplacian and adjacency matrix are equivalent, and the algorithm will sample from both $\ket{a}$ and $\ket{b}$ simultaneously. It also excludes when $N_1$ and $N_2$ are within $o(\sqrt{N})$ of each other, which causes $\gamma_a$ and $\gamma_b$ to be within $o(1/N^{3/2})$ of each other. In this regime, we also have that the Laplacian and adjacency matrix are asymptotically equivalent, and we can use the analysis from the adjacency matrix in the next section.

We also note that $\gamma_a$ and $\gamma_b$ only depend on $N_1$ and $N_2$, which are presumed to be known since the spatial search problem assumes that the graph structure is known. That is, $\gamma_a$ and $\gamma_b$ do not depend on $k_1$ and $k_2$, which may be unknown. For example, say there are $k = 10$ marked vertices. Then the critical $\gamma$'s do not depend on their arrangement, on whether $k_1 = 3$ and $k_2 = 7$, or $k_1 = 6$ and $k_2 = 4$, for instance. This differs from the more complicated ``simplex of complete graphs'' in \cite{Wong9}, where different arrangements of marked vertices can yield different critical $\gamma$'s.

Finally, although the runtimes $t_a$ and $t_b$ do depend on the number of marked vertices $k_1$ and $k_2$ in each vertex set, one can use the same sampling or counting techniques as for Grover's algorithm with an unknown number of marked vertices \cite{Boyer1998}.


\section{Adjacency Walk}

In this section, we now use the adjacency matrix to effect the quantum walk, so the search Hamiltonian is
\[ H = -\gamma A - \!\!\!\!\!\! \sum_{i \in {\rm marked}} \!\!\!\!\! \ketbra{i}{i}. \]
Before analyzing the algorithm, however, we must first define the initial state of the system. Before, when the walk was governed by the Laplacian, the initial state was the equal superposition $\ket{s}$ over all the vertices. With this initial state, the quantum walk alone ($H = -\gamma L$) caused the system to stay in $\ket{s}$, which is expected because without the oracle, no new information is acquired as to where the marked vertices may be.

The equal superposition $\ket{s}$, however, is not an eigenvector of $A$. Then the quantum walk governed by the adjacency matrix alone (\textit{i.e.}, $H = -\gamma A$), will cause $\ket{s}$ to evolve, even though no oracle is being queried and our information is unchanged.

So instead of using $\ket{s}$, we choose the initial state of the system to be
\[ \ket{\sigma} = \frac{1}{\sqrt{2N_1}} \sum_{i \in V_1} \ket{i} + \frac{1}{\sqrt{2N_2}} \sum_{i \in V_2} \ket{i}, \]
which is an eigenstate of $A$ with eigenvalue $\sqrt{N_1N_2}$. With this state, vertices in the left vertex set start with probability $1/2N_1$, and vertices in the right vertex set start with probability $1/2N_2$. Although this is a non-uniform probability distribution over all the vertices, the benefit of starting in $\ket{\sigma}$ is that evolving it by the quantum walk alone (\textit{i.e.}, $H = -\gamma A$) only contributes a global, unobservable phase, so our initial probability distribution is unchanged, as expected since no new information is garnered. Perhaps the most compelling reason to start in $\ket{\sigma}$, however, is that it is the state that naturally evolves by the search algorithm to the marked vertices with certainty; starting in $\ket{s}$, by comparison, yields a worse success probability.

With this initial state and search Hamiltonian, the system evolves in the same 4D subspace shown in Fig.~\ref{fig:bipartite} with orthonormal basis $\{ \ket{a}, \ket{b}, \ket{c}, \ket{d} \}$ as before. In this basis, the initial state is
\[ \ket{\sigma} = \frac{1}{\sqrt{2N_1N_2}} \begin{pmatrix}
	\sqrt{k_1 N_2} \\
	\sqrt{k_2 N_1} \\
	\sqrt{N_2(N_1-k_1)} \\
	\sqrt{N_1(N_2-k_2)} \\
\end{pmatrix}. \]
Using \eqref{eq:A}, the search Hamiltonian is
\begin{equation}
	\label{eq:H_A}
	H = -\gamma \begin{pmatrix}
		\frac{1}{\gamma} & \sqrt{k_1k_2} & 0 & \sqrt{k_1 N_{k2}} \\
		\sqrt{k_1k_2} & \frac{1}{\gamma} & \sqrt{k_2 N_{k1}} & 0 \\
		0 & \sqrt{k_2 N_{k1}} & 0 & \sqrt{N_{k1} N_{k2}} \\
		\sqrt{k_1 N_{k2}} & 0 & \sqrt{N_{k1} N_{k2}} & 0 \\
	\end{pmatrix}.
\end{equation}
where we again denote $N_{ki} = N_i - k_i$.

\begin{figure}
\begin{center}
	\includegraphics{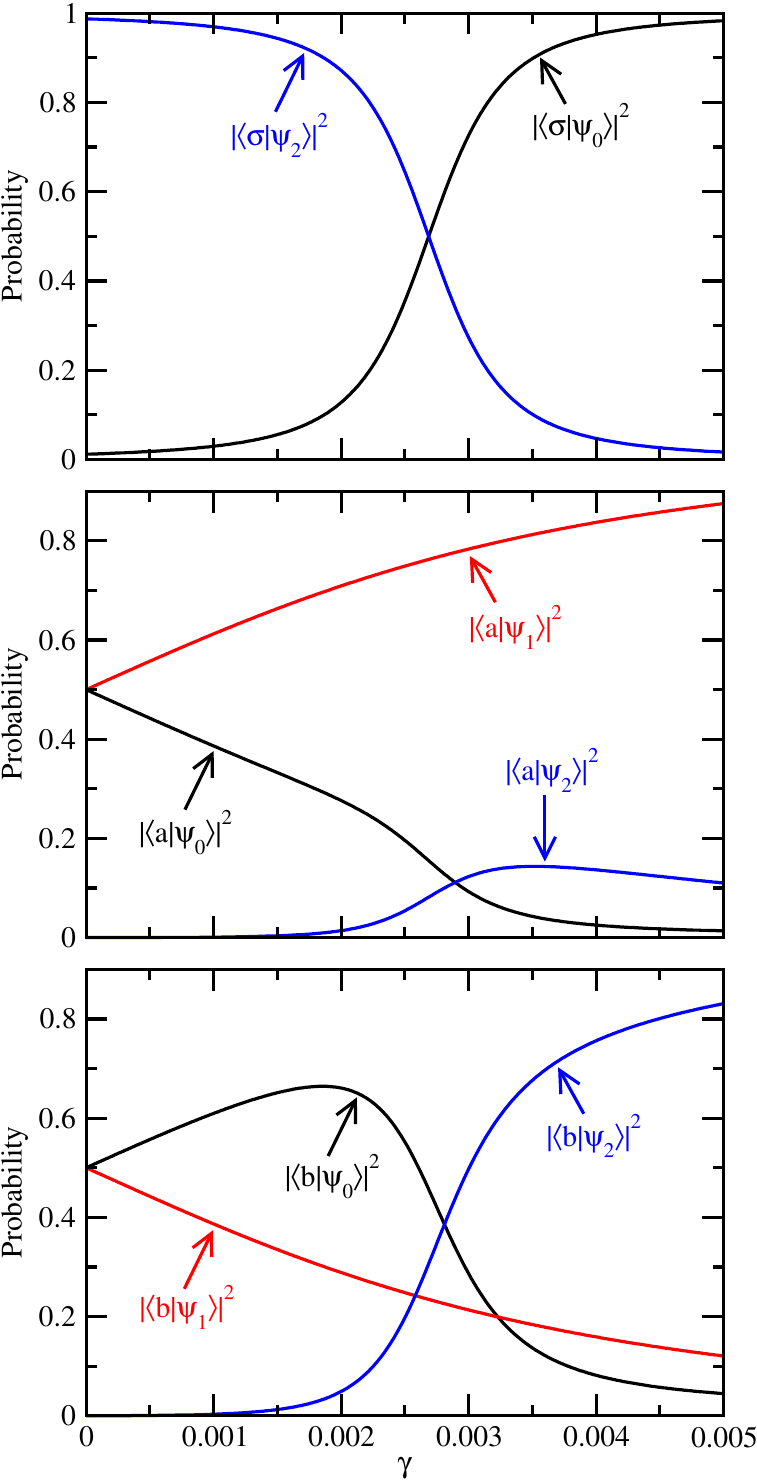}
	\caption{\label{fig:overlap_A} Probability overlaps of $\ket{\sigma}$, $\ket{a}$, and $\ket{b}$ with eigenstates of $H$ for search on the complete bipartite graph with $N_1 = 512$, $N_2 = 256$, $k_1 = 3$, and $k_2 = 5$, where the walk is effected by the adjacency matrix.}
\end{center}
\end{figure}

As before, we can determine how the algorithm depends on $\gamma$ by plotting the probability overlaps of the eigenstates of $H$ with the starting state $\ket{\sigma}$ and the marked vertices $\ket{a}$ and $\ket{b}$, as shown in Fig.~\ref{fig:overlap_A}. When $\gamma$ is small or large, the initial state $\ket{\sigma}$ is approximately the second excited state $\ket{\psi_2}$ or the ground state $\ket{\psi_0}$ of $H$, meaning the system starts in an eigenstate and fails to evolve apart from an unobservable, global phase. But when $\gamma$ takes its critical value of roughly $0.0028$, both $\ket{\psi_0}$ and $\ket{\psi_2}$ are half $\ket{\sigma}$ and half some combination of $\ket{a}$ and $\ket{b}$. Then the system evolves from $\ket{\sigma}$ to that combination of $\ket{a}$ and $\ket{b}$ in time $\pi/\Delta E$, where $\Delta E$ is the energy gap between $\ket{\psi_2}$ and $\ket{\psi_0}$.

\begin{figure}
\begin{center}
	\includegraphics{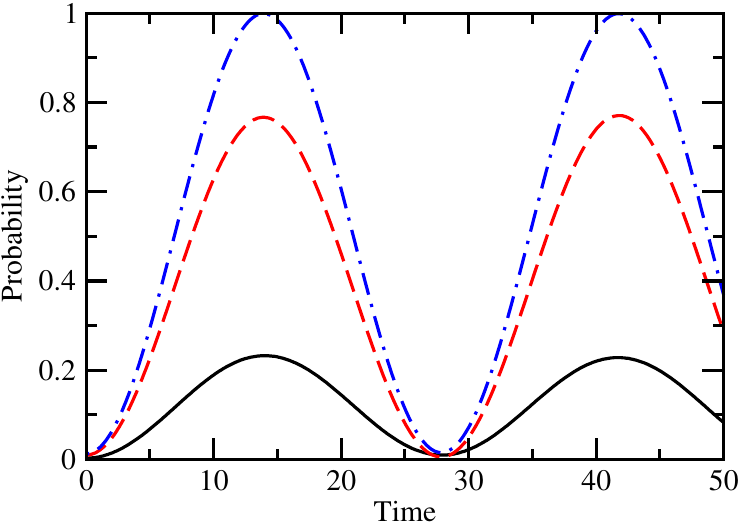}
	\caption{\label{fig:evolution_A_ab} For search on the complete bipartite graph with $N_1 = 512$, $N_2 = 256$, $k_1 = 3$, and $k_2 = 5$, where the walk is effected by the adjacency matrix: the probability in $\ket{a}$ (solid black), the probability in $\ket{b}$ (dashed red), and their sum (dot-dashed blue) when $\gamma = \gamma_*$.}
\end{center}
\end{figure}

Again, proving this simply involves finding the eigenvectors and eigenvalues of the search Hamiltonian \eqref{eq:H_A} when $\gamma$ takes its critical value. As shown in Appendix~\ref{appendix:adjacency} using degenerate perturbation theory \cite{Wong5}, when $\gamma$ takes its critical value of
\[ \gamma_* = \frac{1}{\sqrt{N_1 N_2}} + o\left( \frac{1}{N^{3/2}} \right), \]
then two of the (unnormalized) eigenvectors of the search Hamiltonian \eqref{eq:H_A}, for large $N$, are
\[ \ket{\sigma} \pm \sqrt{\frac{k_1 N_2}{k_2 N_1 + k_1 N_2}} \ket{a} \pm \sqrt{\frac{k_2 N_1}{k_2 N_1 + k_1 N_2}} \ket{b} \]
with corresponding eigenvalues
\[ -1 \mp \sqrt{\frac{k_2 N_1 + k_1 N_2}{2 N_1 N_2}}. \]
So for large $N$, the system evolves from $\ket{\sigma}$ to 
\begin{equation}
	\label{eq:A_final}
	\sqrt{\frac{k_1 N_2}{k_2 N_1 + k_1 N_2}} \ket{a} + \sqrt{\frac{k_2 N_1}{k_2 N_1 + k_1 N_2}} \ket{b},
\end{equation}
in time
\[ t_* = \frac{\pi}{\Delta E} = \frac{\pi \sqrt{N_1 N_2}}{\sqrt{2(k_2 N_1 + k_1 N_2)}}. \]
As a check, in Fig.~\ref{fig:evolution_A_ab}, we plot the probability in $\ket{a}$, $\ket{b}$, and their sum (\textit{i.e.}, the success probability) as the system evolves with $\gamma = \gamma_* = 1/\sqrt{512 \cdot 256} \approx 0.0028$. As expected, at time $t_* = \pi \sqrt{512 \cdot 256} / \sqrt{2(5 \cdot 512 + 3 \cdot 256)} \approx 13.941$, the probability in $\ket{a}$ reaches $3 \cdot 256 / (5 \cdot 512 + 3 \cdot 256) \approx 0.231$, and the probability in $\ket{b}$ reaches $5 \cdot 512 / (5 \cdot 512 + 3 \cdot 256) \approx 0.769$, for a total success probability of $1$.

\begin{table}
	\caption{\label{table:summary} Summary of search on the complete bipartite graph with the quantum walk defined by the Laplacian or adjacency matrix. Note $N_1 \not\approx N_2$ and $N_1 \approx N_2$ denote tolerances of $\omega(\sqrt{N})$ and $o(\sqrt{N})$, and the critical $\gamma$'s have precisions of $o(1/N^{3/2})$.}
	\begin{center}
	\begin{tabular}{ccccc}
		\hline\noalign{\smallskip}
		\textbf{Walk} & \textbf{Critical} $\boldsymbol{\gamma}$ & \textbf{Runtime} & \textbf{Evolution} \\
		\noalign{\smallskip}\hline\noalign{\smallskip}
		$L\ (N_1 \not\approx N_2)$ & $\frac{1}{N_2}$ & $\frac{\pi}{2} \sqrt{\frac{N}{k_1}}$ & $\ket{s} \to \ket{a}$ \\
		$L\ (N_1 \not\approx N_2)$ & $\frac{1}{N_1}$ & $\frac{\pi}{2} \sqrt{\frac{N}{k_2}}$ & $\ket{s} \to \ket{b}$ \\
		$L\ (N_1 \approx N_2)$ or $A$ & $\frac{1}{\sqrt{N_1N_2}}$ & $\frac{\pi \sqrt{N_1 N_2}}{\sqrt{2(k_2 N_1 + k_1 N_2)}}$ & $\ket{\sigma} \to \!\! \begin{array}{l} \sqrt{\frac{k_1 N_2}{k_2 N_1 + k_1 N_2}} \ket{a} \\ \enspace + \sqrt{\frac{k_2 N_1}{k_2 N_1 + k_1 N_2}} \ket{b} \end{array}$ \\
		\noalign{\smallskip}\hline
	\end{tabular}
	\end{center}
\end{table}

We stress that this is a qualitatively different behavior from search with the Laplacian. Here the final state is a combination of $\ket{a}$ and $\ket{b}$, so it samples from all the marked vertices, albeit unequally. With the Laplacian in the last section, the system evolved to either $\ket{a}$ or $\ket{b}$ exclusively, depending on the choice of $\gamma$. These behaviors are summarized in Table~\ref{table:summary}.

\begin{figure}
\begin{center}
	\subfloat[]{
		\includegraphics{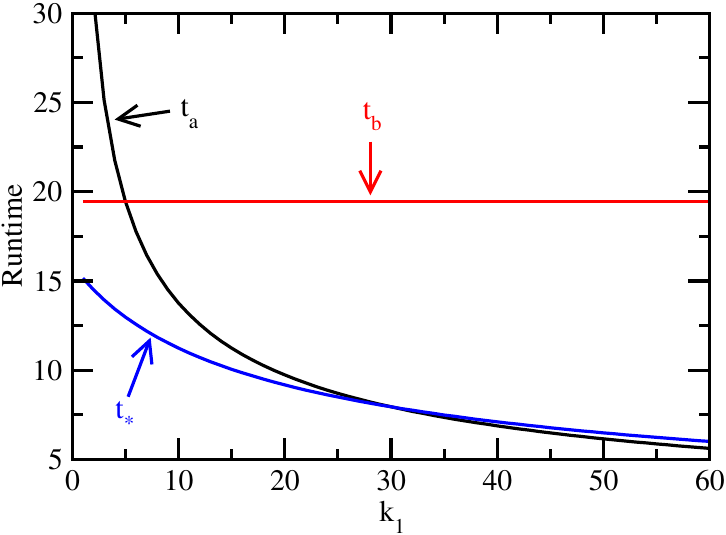}
		\label{fig:compare_N1}
	}

	\subfloat[]{
		\includegraphics{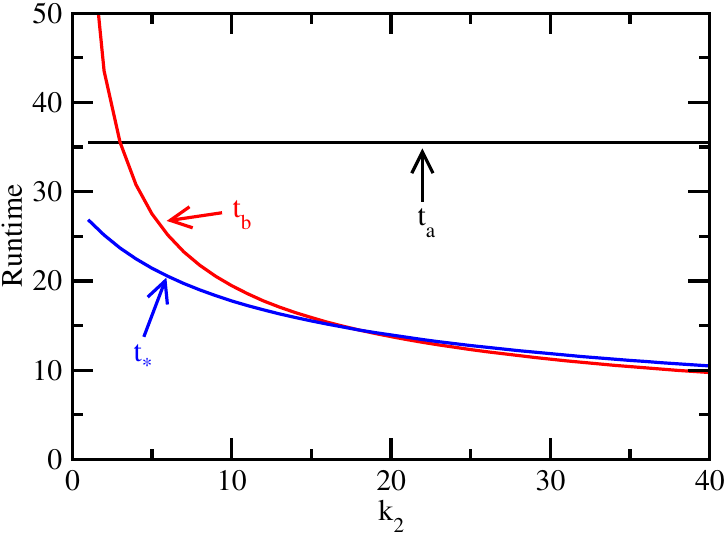}
		\label{fig:compare_N2}
	}
	\caption{Runtimes for search on the complete bipartite graph with (a) $N_1 = 512$, $N_2 = 256$, $k_1$ varying, and $k_2 = 5$, and (b) $N_1 = 512$, $N_2 = 1024$, $k_1 = 3$, and $k_2$ varying.}
\end{center}
\end{figure}

Except for the special case when $N_1 \approx N_2$, which causes the graph to be approximately regular and the Laplacian and adjacency matrix to generate asymptotically equivalent walks, the search algorithms generally yield different runtimes. By comparing $t_a$, $t_b$, and $t_*$, we find that when $N_1 > N_2$, the adjacency walk's runtime $t_*$ is always faster than the Laplacian's $t_b$, and it is also faster than the Laplacian's $t_a$ when
\[ k_1 < k_2 \frac{N_1}{N_2} \frac{N_1 + N_2}{N_1 - N_2}. \]
For example, with $N_1 = 512$, $N_2 = 256$, and $k_2 = 5$, we find that $t_*$ is faster than $t_a$ when $k_1 < 30$, as verified in Fig.~\ref{fig:compare_N1}. Similarly, when $N_1 < N_2$, then $t_*$ is always faster than $t_a$, and it is also faster than $t_b$ when
\[ k_2 < k_1 \frac{N_2}{N_1} \frac{N_1 + N_2}{N_2 - N_1}. \]
For example, with $N_1 = 512$, $N_2 = 1024$, and $k_1 = 3$, we find that $t_*$ is faster than $t_b$ when $k_2 < 18$, as verified in Fig.~\ref{fig:compare_N2}. These results indicating when the Laplacian or adjacency matrix searches faster are summarized in Table~\ref{table:faster}.

\begin{table}
	\caption{\label{table:faster} Summary of when search on the complete bipartite graph is faster using the Laplacian or adjacency matrix, assuming $N_1$ and $N_2$ are $\omega(\sqrt{N})$ of each other so that the quantum walks are different.}
	\begin{center}
	\begin{tabular}{cccc}
		\hline\noalign{\smallskip}
		\multicolumn{2}{c}{\textbf{Conditions}} & $\boldsymbol{L}$ \textbf{or} $\boldsymbol{A}$ \textbf{Faster} \\
		\noalign{\smallskip}\hline\noalign{\smallskip}
		$N_1 > N_2$ & $k_1 < k_2 \frac{N_1}{N_2} \frac{N_1 + N_2}{N_1 - N_2}$ & $A$ Faster \\[1ex]
		& $k_1 > k_2 \frac{N_1}{N_2} \frac{N_1 + N_2}{N_1 - N_2}$ & $L\ (t_a)$ Faster \\[1ex]
		& $k_1 = k_2 \frac{N_1}{N_2} \frac{N_1 + N_2}{N_1 - N_2}$ & $A$ \& $L\ (t_a)$ Same \\
		\noalign{\smallskip}\hline\noalign{\smallskip}
		$N_1 < N_2$ & $k_2 < k_1 \frac{N_2}{N_1} \frac{N_1 + N_2}{N_2 - N_1}$ & $A$ Faster \\[1ex]
		& $k_2 > k_1 \frac{N_2}{N_1} \frac{N_1 + N_2}{N_2 - N_1}$ & $L\ (t_b)$ Faster \\[1ex]
		 & $k_2 = k_1 \frac{N_2}{N_1} \frac{N_1 + N_2}{N_2 - N_1}$ & $A$ \& $L\ (t_b)$ Same \\
		\noalign{\smallskip}\hline
	\end{tabular}
	\end{center}
\end{table}

Another way to compare the two quantum walk search algorithms is by considering a slightly different problem. Rather than searching for one of $k$ marked vertices, say we instead want to find all $k$ marked vertices. With the graph Laplacian, we can choose $\gamma = \gamma_a = 1/N_2$ so that the system evolves to $\ket{a}$, which is a uniform superposition over the $k_1$ marked vertices in the left vertex set. Classically, sampling all $k_1$ of these marked vertices is simply the ``coupon collector's problem'' from classical probability theory \cite{MR1995}, and the expected number of repetitions is $k_1 H_{k_1}$, where $H_n = 1 + 1/2 + 1/3 + \dots + 1/n = \Theta(\ln(n))$ denotes the $n$-th harmonic number. Similarly, with the graph Laplacian and $\gamma = \gamma_b = 1/N_1$, the system evolves to $\ket{b}$, and we expect to make $k_2 H_{k_2}$ repetitions of the algorithm to find all $k_2$ marked vertices in the right vertex set. Thus with the Laplacian, we expect to run the algorithm $k_1 H_{k_1} + k_2 H_{k_2}$ times to find all $k$ marked vertices. As an example, if $k_1 = 3$ and $k_2 = 5$, this yields $3H_3 + 5H_5 = 203/12 \approx 16.917$ repetitions.

By contrast, searching with the adjacency matrix results in a non-uniform final state \eqref{eq:A_final}. To sample all marked vertices from this final state, we intuitively need more repetitions than the Laplacian case because there are now a greater number of vertices from which we might repeatedly sample. Mathematically, this is the coupon collector's problem generalized to non-uniform probabilities \cite{vonSchelling1934,vonSchelling1954,Flajolet1992}, which has expected value
\[ \int_0^\infty \left( 1 - \left( 1 - e^{\frac{-N_2 t}{k_2 N_1 + k_1 N_2}} \right)^{k_1} \left( 1 - e^{\frac{-N_1 t}{k_2 N_1 + k_1 N_2}} \right)^{k_2} \right) dt. \]
For example, with $N_1 = 512$, $N_2 = 256$, $k_1 = 3$, and $k_2 = 5$, this numerically integrates to 26.368, which is greater than the Laplacian's 16.917. Thus to find all marked vertices, the Laplacian is expected to be faster.

We end by commenting on the success of the algorithm if the initial state is the equal superposition $\ket{s}$ over the vertices, rather than the state $\ket{\sigma}$ that evolves fully to $\ket{a}$ and $\ket{b}$. Consider the state
\[ \ket{\delta} = \frac{1}{\sqrt{2N_1N_2}} \begin{pmatrix}
	-\sqrt{k_1 N_2} \\
	\sqrt{k_2 N_1} \\
	-\sqrt{N_2(N_1-k_1)} \\
	\sqrt{N_1(N_2-k_2)} \\
\end{pmatrix}, \]
which is orthonormal to $\ket{\sigma}$. The equal superposition $\ket{s}$ can be expressed as a linear combination of $\ket{\sigma}$ and $\ket{\delta}$:
\[ \ket{s} = \frac{1}{\sqrt{2N}} \! \left[ \! \left( \sqrt{N_1} + \sqrt{N_2} \right) \ket{\sigma} + \left( -\sqrt{N_1} + \sqrt{N_2} \right) \ket{\delta} \right] \!. \]
As shown in Appendix~\ref{appendix:adjacency}, $\ket{\delta}$ is approximately an eigenstate of the search Hamiltonian \eqref{eq:H_A}, and so it does not evolve apart from a global, unobservable phase. Thus if we start in $\ket{s}$, the part in $\ket{\sigma}$ evolves to a combination of $\ket{a}$ and $\ket{b}$. Meanwhile, the part in $\ket{\delta}$ only acquires a phase, but since it has negligible projections onto $\ket{a}$ and $\ket{b}$ compared to its components in $\ket{c}$ and $\ket{d}$, it has negligible affect on the success probability. Thus at the runtime of $t_*$, the system will reach a success probability of roughly
\[ \left( \frac{\sqrt{N_1} + \sqrt{N_2}}{\sqrt{2N}} \right)^2 = \frac{1}{2} + \frac{\sqrt{N_1 N_2}}{N} > \frac{1}{2} \]
from the $\ket{\sigma}$ piece. Since this is lower-bounded by $1/2$, if we start in $\ket{s}$, up to two repetitions of the algorithm are expected, on average, to find a marked vertex. To emphasize again, this contrasts with starting in $\ket{\sigma}$, which naturally evolves to the marked vertices with certainty.


\section{Conclusion}

The continuous-time quantum walk can be defined in a variety of ways, so long as it is generated by a Hermitian operator that respects the locality of the graph. The two most common definitions utilize the graph Laplacian $L$ and adjacency matrix $A$. Although these are equivalent for regular graphs, they differ for non-regular graphs.

In this paper, we investigated how each type of walk differs when solving the spatial search problem on the complete bipartite graph with multiple marked vertices, which in general is non-regular. This leads to qualitative and quantitative differences between the two walks. For the Laplacian walk, two critical jumping rates $\gamma_a$ and $\gamma_b$ exist, which respectively cause the system to evolve from the equal superposition over the vertices $\ket{s}$ to either $\ket{a}$ or $\ket{b}$, the marked vertices in each vertex set. This contrasts with the adjacency walk, which utilizes a different starting state $\ket{\sigma}$, and a single critical jumping rate $\gamma_*$ causes the system to evolve to a combination of $\ket{a}$ and $\ket{b}$. Besides these qualitative differences, the runtimes of the respective algorithms are, in general, different, and depending on the number of marked and unmarked vertices in each vertex set, one walk can outperform the other. Thus the choice of the Laplacian or adjacency matrix to effect the walk has important algorithmic consequences.


\appendix

\section{\label{appendix:Laplacian} Eigensystem for Laplacian Walk}

The search Hamiltonian, when walking with the Laplacian, is given in \eqref{eq:H}. Finding the exact eigenvectors and eigenvalues of $H$ is intractable, but they can be approximated for large $N$ using degenerate perturbation theory, which also gives a way to find the critical $\gamma$'s \cite{Wong5}. To do this, we separate the Hamiltonian \eqref{eq:H} into its leading- and higher-order terms, \textit{i.e.}, $H = H^{(0)} + H^{(1)} + \dots $, where
\begin{gather*}
	H^{(0)} = -\gamma \begin{pmatrix}
		\frac{1}{\gamma} - N_2 & 0 & 0 & 0 \\
		0 & \frac{1}{\gamma} - N_1 & 0 & 0 \\
		0 & 0 & -N_2 & \sqrt{N_1 N_2} \\
		0 & 0 & \sqrt{N_1 N_2} & -N_1 \\
	\end{pmatrix}, \\
	H^{(1)} = -\gamma \begin{pmatrix}
		0 & 0 & 0 & \sqrt{k_1N_2} \\
		0 & 0 & \sqrt{k_2N_1} & 0 \\
		0 & \sqrt{k_2N_1} & 0 & 0 \\
		\sqrt{k_1N_2} & 0 & 0 & 0 \\
	\end{pmatrix}.
\end{gather*}
The idea is to first find the eigenvectors and eigenvalues of the leading-order Hamiltonian $H^{(0)}$, which is a much simpler matrix. Then we add the next-order corrections $H^{(1)}$ and see how this perturbation modifies them. This gives an approximation for the eigenvectors and eigenvalues of $H$.

To begin, the eigenvectors and eigenvalues of $H^{(0)}$ are
\begin{gather*}
	\ket{a}, \quad \gamma N_2 - 1 \\
	\ket{b}, \quad \gamma N_1 - 1 \\
	\ket{r} = \frac{1}{\sqrt{N}} \left( \sqrt{N_1} \ket{c} + \sqrt{N_2} \ket{d} \right), \quad 0 \\
	\frac{1}{\sqrt{N}} \left( -\sqrt{N_2} \ket{c} + \sqrt{N_1} \ket{d} \right), \quad \gamma(N_1 + N_2).
\end{gather*}
Note that the third eigenvector, which we call $\ket{r}$, is approximately $\ket{s}$ for large $N$ since the $k_1$ and $k_2$ terms in $\ket{s}$ are dominated by $N_1$ and $N_2$ for large $N$.

Now we include the perturbation $H^{(1)}$ to see how these leading-order eigenvectors and eigenvalues change. If they are non-degenerate, then $H^{(1)}$ can only contribute higher-order terms to the eigenvectors and eigenvalues, so the starting state $\ket{s} \approx \ket{r}$ is still an approximate eigenvector of the perturbed system for large $N$ \cite{Griffiths2005}. If the leading-order eigenvectors are degenerate, however, the behavior is vastly different. For example, say
\[ \gamma_a = \frac{1}{N_2} \]
so that $\ket{a}$ and $\ket{r}$ are degenerate to leading-order. Then the addition of $H^{(1)}$ causes two linear combinations of them
\[ \alpha_a \ket{a} + \alpha_r \ket{r} \]
become eigenstates of the perturbed system \cite{Griffiths2005}. To find the coefficients, we solve the eigenvalue problem
\[ \begin{pmatrix}
	H_{aa} & H_{ar} \\
	H_{ra} & H_{rr} \\
\end{pmatrix} \begin{pmatrix}
	\alpha_a \\
	\alpha_r \\
\end{pmatrix} = E \begin{pmatrix}
	\alpha_a \\
	\alpha_r \\
\end{pmatrix}, \]
where $H_{ar} = \langle a | H^{(0)} + H^{(1)} | r \rangle$, etc. Evaluating the matrix elements,
\[ \begin{pmatrix}
	0 & -\sqrt{\frac{k_1}{N}} \\
	-\sqrt{\frac{k_1}{N}} & 0 \\
\end{pmatrix} \begin{pmatrix}
	\alpha_a \\
	\alpha_r \\
\end{pmatrix} = E \begin{pmatrix}
	\alpha_a \\
	\alpha_r \\
\end{pmatrix}. \]
This has solutions
\[ \frac{1}{\sqrt{2}} \left( \pm \ket{a} + \ket{r} \right), \quad E = \mp \sqrt{\frac{k_1}{N}}. \]
Since $\ket{r} \approx \ket{s}$, for large $N$, the (unnormalized) eigenvectors and eigenvalues of $H$ when $\gamma = \gamma_a$ are
\[ \ket{s} \pm \ket{a}, \quad E = \mp \sqrt{\frac{k_1}{N}}, \]
as stated in the main text of the paper.

Note that $\ket{s} + \ket{a}$ and $\ket{s} - \ket{a}$ are non-degenerate eigenstates of the perturbed system, so the first-order correction $H^{(1)}$ has lifted the degeneracy. Then any higher-order corrections ($H^{(2)}$, $H^{(3)}$, etc.) will not significantly change these eigenstates and eigenvalues, meaning any contributions from them will go to zero more quickly than the terms we have derived \cite{Griffiths2005}, and hence we can safely ignore them.

Using the method of Section IV of \cite{Wong16}, we can find how close $\gamma$ must be to its critical value. Say $\gamma$ is within $\epsilon$ of its critical value, \textit{i.e.}, $\gamma = \gamma_a + \epsilon = 1/N_2 + \epsilon$. If $\epsilon$ is small such that $\ket{a}$ and $\ket{r}$ are near-degenerate to leading-order, then the perturbation still causes $\alpha_a \ket{a} + \alpha_r \ket{r}$ to be eigenvectors of the perturbed system. To find the coefficients, one solves an eigenvalue problem, which has a leading-order term in $\epsilon$ scaling as $\epsilon N_2$ due to the component $H_{aa} = \langle a | H^{(0)} + H^{(1)} | a \rangle$. Thus for the system to retain the error-free energy gap of $\Theta(1/\sqrt{N})$, we need $\epsilon N_2 = o(1/\sqrt{N})$. That is, $\epsilon = o(1/N^{3/2})$, which is the precision stated in the main text of the paper.

Similarly, if
\[ \gamma_b = \frac{1}{N_1} \]
so that $\ket{b}$ and $\ket{r}$ are degenerate to leading-order, then the perturbation $H^{(1)}$ causes two linear combinations $\alpha_b \ket{b} + \alpha_r \ket{r}$ to be eigenvectors of the perturbed system. Solving a similar eigenvalue problem for the coefficients $\alpha_b$ and $\alpha_r$, we find that the perturbed eigenstates are
\[ \frac{1}{\sqrt{2}} \left( \pm \ket{b} + \ket{r} \right), \quad E = \mp \sqrt{\frac{k_2}{N}}. \]
Since $\ket{r} \approx \ket{s}$, for large $N$, the (unnormalized) eigenvectors and eigenvalues of $H$ when $\gamma = \gamma_b$ are
\[ \ket{s} \pm \ket{b}, \quad E = \mp \sqrt{\frac{k_2}{N}}, \]
as stated in the main text of the paper.

For the precision with which $\gamma$ must be chosen to its critical value, we can again use the argument of \cite{Wong16} to show that $\gamma$ must be within $o(1/N^{3/2})$ of $\gamma_b$, as stated in the main text of the paper.

\begin{figure}
\begin{center}
	\subfloat[]{
		\includegraphics{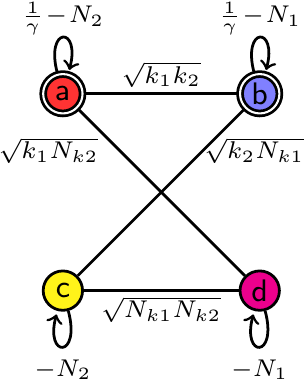}
		\label{fig:diagram_L_H}
	} \quad \quad \quad
	\subfloat[]{
		\includegraphics{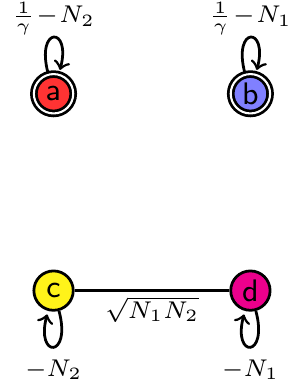}
		\label{fig:diagram_L_H0}
	}
	\caption{Apart from a factor of $-\gamma$, the (a) Hamiltonian for search on the complete bipartite graph, where the walk is effected by the Laplacian, and (b) the leading-order terms. Note $N_{ki} = N_i - k_i$.}
\end{center}
\end{figure}

This calculation using degenerate perturbation theory can also be understood diagrammatically \cite{Wong8}. The search Hamiltonian \eqref{eq:H} can be depicted as shown in Fig.~\ref{fig:diagram_L_H}. Keeping only the leading-order terms, $H^{(0)}$ is shown in Fig.~\ref{fig:diagram_L_H0}, and the diagram reveals its four eigenvectors: $\ket{a}$, $\ket{b}$, and two linear combinations of $\ket{c}$ and $\ket{d}$ (one of which we called $\ket{r}$). By choosing $\ket{a}$ to be degenerate with $\ket{r}$, the perturbation, which restores the missing edges, causes $\ket{a}$ and $\ket{r}$ to mix. Similarly, if $\ket{b}$ is degenerate with $\ket{r}$, the perturbation causes them to mix.


\section{\label{appendix:adjacency} Eigensystem for Adjacency Walk}

Similar to the last section, we approximate the eigenvectors and eigenvalues of the search Hamiltonian \eqref{eq:H_A} using degenerate perturbation theory \cite{Wong5}. Breaking $H$ into its leading- and higher-order terms,
\begin{gather*}
	H^{(0)} = -\gamma \begin{pmatrix}
		\frac{1}{\gamma} & 0 & 0 & 0 \\
		0 & \frac{1}{\gamma} & 0 & 0 \\
		0 & 0 & 0 & \sqrt{N_1 N_2} \\
		0 & 0 & \sqrt{N_1 N_2} & 0 \\
	\end{pmatrix}, \\
	H^{(1)} = -\gamma \begin{pmatrix}
		0 & 0 & 0 & \sqrt{k_1N_2} \\
		0 & 0 & \sqrt{k_2N_1} & 0 \\
		0 & \sqrt{k_2N_1} & 0 & 0 \\
		\sqrt{k_1N_2} & 0 & 0 & 0 \\
	\end{pmatrix}.
\end{gather*}
$H^{(0)}$ has eigenvectors and eigenvalues
\begin{gather*}
	\ket{a}, \quad -1 \\
	\ket{b}, \quad -1 \\
	\ket{u} = \frac{1}{\sqrt{2}} \left( \ket{c} + \ket{d} \right), \quad -\gamma \sqrt{N_1 N_2} \\
	\ket{v} = \frac{1}{\sqrt{2}} \left( -\ket{c} + \ket{d} \right), \quad \gamma \sqrt{N_1 N_2}.
\end{gather*}
Note that the third eigenvector, which we call $\ket{u}$, is approximately $\ket{\sigma}$ for large $N$. Also, the fourth eigenvector, which we call $\ket{v}$, is approximately $\ket{\delta}$ for large $N$.

\begin{figure}
\begin{center}
	\subfloat[]{
		\includegraphics{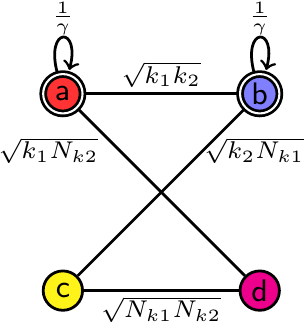}
		\label{fig:diagram_A_H}
	} \quad \quad \quad
	\subfloat[]{
		\includegraphics{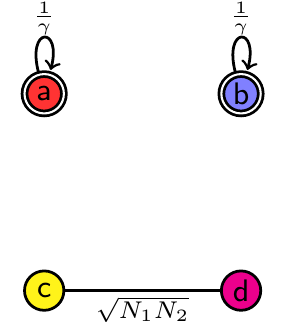}
		\label{fig:diagram_A_H0}
	}
	\caption{Apart from a factor of $-\gamma$, the (a) Hamiltonian for search on the complete bipartite graph, where the walk is effected by the adjacency matrix, and (b) the leading-order terms. Note $N_{ki} = N_i - k_i$.}
\end{center}
\end{figure}

Now we include the perturbation $H^{(1)}$. If $\ket{u}$ is non-degenerate to leading-order, then the starting state $\ket{\sigma} \approx \ket{u}$ approximately remains an eigenstate of the perturbed system. On the other hand, if
\[ \gamma_* = \frac{1}{\sqrt{N_1 N_2}}, \]
then $\ket{a}$, $\ket{b}$, and $\ket{u}$ are degenerate to leading-order, and the three linear combinations of them
\[ \alpha_a \ket{a} + \alpha_b \ket{b} + \alpha_u \ket{u} \]
will be eigenvectors of the perturbed system. To find the coefficients, we solve the eigenvalue problem
\[ \begin{pmatrix}
	H_{aa} & H_{ab} & H_{au} \\
	H_{ba} & H_{bb} & H_{bu} \\
	H_{ua} & H_{ub} & H_{uu} \\
\end{pmatrix} \begin{pmatrix}
	\alpha_a \\
	\alpha_b \\
	\alpha_u \\
\end{pmatrix} = E \begin{pmatrix}
	\alpha_a \\
	\alpha_b \\
	\alpha_u \\
\end{pmatrix}, \]
where $H_{ab} = \langle a | H^{(0)} + H^{(1)} | b \rangle$, etc. 
Evaluating the matrix components, we get
\[ \begin{pmatrix}
	-1 & 0 & -\sqrt{\frac{k_1}{2 N_1}} \\
	0 & -1 & -\sqrt{\frac{k_2}{2 N_2}} \\
	-\sqrt{\frac{k_1}{2 N_1}} & -\sqrt{\frac{k_2}{2 N_2}} & -1 \\
\end{pmatrix} \begin{pmatrix}
	\alpha_a \\
	\alpha_b \\
	\alpha_u \\
\end{pmatrix} = E \begin{pmatrix}
	\alpha_a \\
	\alpha_b \\
	\alpha_u \\
\end{pmatrix}. \]
Solving this yields (unnormalized) eigenstates
\begin{gather*}
	\psi_{-1} = -\sqrt{\frac{k_2 N_1}{k_1 N_2}} \ket{a} + \ket{b} \\
	\psi_- = \sqrt{\frac{k_1 N_2}{k_2 N_1 + k_1 N_2}} \ket{a} + \sqrt{\frac{k_2 N_1}{k_2 N_1 + k_1 N_2}} \ket{b} + \ket{u} \\
	\psi_+ = -\sqrt{\frac{k_1 N_2}{k_2 N_1 + k_1 N_2}} \ket{a} - \sqrt{\frac{k_2 N_1}{k_2 N_1 + k_1 N_2}} \ket{b} + \ket{u}
\end{gather*}
with corresponding eigenvalues
\begin{gather*}
	E_{-1} = -1 \\
	E_- = -1 - \sqrt{\frac{k_2 N_1 + k_1 N_2}{2 N_1 N_2}} \\
	E_+ = -1 + \sqrt{\frac{k_2 N_1 + k_1 N_2}{2 N_1 N_2}}.
\end{gather*}
Note that $\psi_\mp$ and its corresponding eigenvalues $E_\mp$ were stated in the main text of the paper, with $\ket{u}$ replaced by $\ket{\sigma}$, assuming large $N$.

As before, we can use the argument of \cite{Wong16} to find the precision with which $\gamma$ must be chosen to its critical value. If $\gamma$ is within $\epsilon$ of $\gamma_*$, then there is a term scaling as $\epsilon\sqrt{N_1N_2}$ that appears in the perturbative calculation, which is leading-order in $\epsilon$. For this to be small enough to not interfere with the energy gap of $\Theta(1/\sqrt{N})$, we get $\epsilon = o(1/N^{3/2})$, which is the precision stated in the main text of the paper.

The Hamiltonian \eqref{eq:H_A} can be represented diagrammatically \cite{Wong8}, as depicted in Fig.~\ref{fig:diagram_A_H}. Keeping only the leading-order terms, $H^{(0)}$ is shown in Fig.~\ref{fig:diagram_A_H0}, and we see that $\ket{a}$ and $\ket{b}$ are always degenerate since they have the same self-loop. Making these degenerate with $\ket{u}$, which is a combination of $\ket{c}$ and $\ket{d}$, the perturbation restores the missing edges and mixes $\ket{a}$, $\ket{b}$, and $\ket{u}$.

Finally, since $\gamma > 0$, the leading-order eigenvector $\ket{v}$ is never degenerate with the others. So it remains an approximate eigenstate of the perturbed system. Since $\ket{v} \approx \ket{\delta}$, we get that $\ket{\delta}$ is approximately an eigenvector of $H$ for large $N$, as stated in the main text of the paper.


\begin{acknowledgements}
	TW and NN were supported by the European Union Seventh Framework Programme (FP7/2007-2013) under the QALGO (Grant Agreement No.~600700) project, and the ERC Advanced Grant MQC. LT was supported by CNPq CSF/BJT grant reference 301181/2014-4.
\end{acknowledgements}


\bibliographystyle{qinp}
\bibliography{refs}

\end{document}